\documentclass[twocolumn]{aastex63}

\newcommand{\Msun}{\ensuremath{{\rm M}_\odot}}
\newcommand{\Lsun}{\ensuremath{{\rm L}_\odot}}
\newcommand{\kms}{\ensuremath{\,{\rm km\,s}^{-1}}}

\begin{document}

\title{K-Band Imaging of the Nearby, Clumpy Turbulent Disk Galaxy DYNAMO G04-1}

\author[0000-0001-7013-6921]{Heidi A. White}
\affiliation{Department of Astronomy and Astrophysics, University of Toronto, 50 St. George St., Toronto, ON, M5S 3H8, Canada}
\affiliation{Dunlap Institute for Astronomy and Astrophysics, University of Toronto, 50 St. George St., Toronto, ON, M5S 3H8, Canada}
\author[0000-0003-0645-5260]{Deanne B. Fisher}
\affiliation{Centre for Astrophysics and Supercomputing, Swinburne University of Technology, P.O. Box 218, Hawthorn, VIC 3122, Australia}
\affiliation{ARC Centre of Excellence for All Sky Astrophysics in 3 Dimensions (ASTRO 3D), Australia}
\author{Roberto G. Abraham}
\affiliation{Department of Astronomy and Astrophysics, University of Toronto, 50 St. George St., Toronto, ON, M5S 3H8, Canada}
\affiliation{Dunlap Institute for Astronomy and Astrophysics, University of Toronto, 50 St. George St., Toronto, ON, M5S 3H8, Canada}
\author{Karl Glazebrook}
\affiliation{Centre for Astrophysics and Supercomputing, Swinburne University of Technology, P.O. Box 218, Hawthorn, VIC 3122, Australia}
\affiliation{ARC Centre of Excellence for All Sky Astrophysics in 3 Dimensions (ASTRO 3D), Australia}
\author{Danail Obreschkow}
\affiliation{International Centre for Radio Astronomy Research (ICRAR), University of Western Australia, M468, Crawley, WA 6009, Australia}


\begin{abstract}
We present a case study of stellar clumps in G04-1, a clumpy, turbulent disk galaxy located at $z$ = 0.13 from the DYNAMO sample, using adaptive optics enabled K-band imaging ($\sim2.25$ kpc/arcsec) with Keck/NIRC2. We identify 15 stellar clumps in G04-1 with a range of masses from $3.6 \times 10^{6}$ to $2.7\times 10^{8}\ \Msun$, and with a median mass of $\sim2.9\times 10^{7}\ \Msun$. Note that these masses decrease by about one-half when we apply a light correction for the underlying stellar disk. A majority (12 of 15) of clumps observed in the $K_{P}$-band imaging have associated components in H$\alpha$ maps ($\sim2.75$ kpc/arcsec; $<$R$_{clump}>\ \sim$500 pc) and appear co-located ($\overline{\Delta x} \sim 0.1 \arcsec$). Using \textit{Hubble Space Telescope} WFC/ACS observations with the F336W and F467M filters, we also find evidence of radial trends in clump stellar properties: clumps closer to the centre of G04-1 are more massive (consistent with observations at high-$z$) and appear more red, suggesting they may be more evolved. Using our high-resolution data, we construct a star forming main sequence for G04-1 in terms of spatially-resolved quantities and find that all regions (both clump and intra-clump) within the galaxy are experiencing an enhanced mode of star formation routinely observed in galaxies at high-$z$. In comparison to recent simulations, our observations of a number of clumps with masses $10^{7}-10^{8}\ \Msun$ is not consistent with strong radiative feedback in this galaxy.
\end{abstract}

\section{Introduction}\label{sec:intro}
The morphology of star forming galaxies at $1\leq z \leq3$ is often dominated by giant ($\sim$1 kpc) and massive ($\gtrsim$10$^{8-9}\ \Msun$) clumps which contribute a significant fraction ($\sim$20-30\%) of the total system's star formation. These clumpy structures are consistently observed via tracers of young stars, e.g. in the rest-UV and -optical (\citealt{cowie1995,elmegreen2004,elmegreen2005,elmegreen2007,guo2012}) and in ionized gas observations \citep{genzel2008}. This irregular morphology was initially attributed to galaxy merger activity and some work examining $z>1$ samples has indicated gravitational interactions can drive clumpiness in galaxies (e.g. \citealt{puech2010,calabro2019}). However, resolved kinematic studies (e.g. \citealt{genzel2006,fs2006,fs2008,fs2009,shapiro2008,epinat2012}) have shown that a large fraction of systems ($>60\%$) do not appear to be mergers. Instead, such systems are often identified as orderly rotators, with clumps thought to form in situ via Toomre instabilities in a gas-rich ($f_{gas}\sim$50-60\%) disk \citep{tacconi2020}.

There is a long standing concept, motivated largely by simulations \citep{noguchi1999,elmegreen2008,immeli2004a,immeli2004b,dekel2009} that these observed clumps could contribute significantly to the formation of bulges in disk galaxies. The processes which allow for this migration (e.g. decreasing angular momentum due to dynamical friction) require clump in-spiral timescales of order a few 10$^{8}$ yr. However, the degree to which clumps can contribute to bulge growth via this mechanism depends on their viability, e.g. whether or not they are disrupted by their own stellar feedback within a few orbital times. Observations of normal $1<z<3$ star forming galaxies observe a range of lifetimes for clumps (100 - 650 Myr; \citealt{zanella2019,fs2020}) suggest that some appear sufficiently long-lived to migrate. This is typically the case for higher mass clumps with $M_{\star} > 10^{9}\Msun$.
If clumps survive long enough to in-spiral, one should expect a radial gradient within the galaxy, in particular with age. Evidence for a radial gradient of clump properties is seen in observations (such as in SINS and CANDELS galaxies; e.g. \citealt{fs2011,fs2020,guo2012,ginzburg2021}) and these observations appear consistent with simulations \citep{bournaud2014,mandelker2017,dekel2021} where clumps closer to the galaxy nucleus appear redder, older, denser, and more massive. Other studies at high-$z$ (e.g. \citealt{guo2018,fs2011,zanella2019}) report little or only slight evidence of a trend between age and galactocentric radius, although small number statistics may influence some of these findings. However, there may be some ambiguity in using radial trends as evidence for long-lived clumps as radial trends are also predicted as a result of inside-out disk formation \citep{murray2010,genel2012}.
More recent theoretical work suggests that though clumps may play a role in bulge formation, significant contributions from large gaseous inflows within these disks is likely very important for bulge formation as well \citep{db2014,zolotov2015}. Moreover, galaxies at $z=1-3$ likely experience multiple periods of instability, and one set of clumps need not explain the full mass of bulges at $z=0$ \citep{tacchella2016}.

As the star formation rate densities of massive clumps are observed to be very high, stellar feedback effects are predicted to play a crucial role in determining the fate of these clumps and how they contribute to their host galaxy (i.e. in building bulges). Consequently, the stellar masses of clumps are both a strong indicator of their viability and an important clue into the dominant forms of feedback at play. Observations of clumps in $1<z<3$ galaxies suggest they are characteristically massive, upwards of $10^{8-9} \Msun$ \citep{elmegreen2007,guo2012,soto2017}. The true sizes and masses of these clumps remain uncertain, however due to the inherent resolution limits imposed on observations at high-$z$. Work probing different spatial scales (for example $\sim 1$ kpc as in \citealt{fs2011} to a few hundred pc \citealt{livermore2012,livermore2015}) consistently find derived clump properties are resolution-dependent and vary with spatial scale. This is supported by investigations of the resolution dependence of clump mass using cosmological simulations (e.g. \citealt{huertas2020,meng2020}). \citet{dz2017} examined clumps in both lensed and unlensed star-forming galaxies at $1.1 < z < 3.6$ and find masses ranging $10^{7} - 10^{9}\Msun$. They also show that the limited resolution available in observations of non-lensed systems works along side clump clustering and projection effects to artificially enhance derived mass estimates. This is also supported by some studies at lower redshift: \citet{larson2020} study clumpy star formation in $z<0.1$ luminous infrared galaxies (LIRGs) and find clumps are overall much less massive ($\sim 10^{5}\Msun$).

Results from numerical work indicate that the details of feedback models have a significant impact on the resulting stellar masses of clumps \citep{tamburello2015,mayer2016,mandelker2017,dekel2021}. Studies where feedback effects are primarily modeled via radiation pressure and supernovae (e.g. \citealt{bournaud2007,bournaud2014}) routinely find long-lived clumps which migrate inward and contribute to bulge growth. \cite{mandelker2017} find that including radiation pressure increases the surface density and mass thresholds for clumps remaining undisrupted for a few free-fall times. Alternatively, \cite{hopkins2012} uses a feedback recipe that includes radiation pressure and high photon trapping fractions. These combined effects produce star forming clumps which are short-lived ($10-100$ Myr) and rapidly disrupted, transforming $\sim$10\% of their gas into stars \citep{hopkins2012}. Indeed, Hopkins et al. predict that the distribution of older stellar populations within clumpy star forming galaxies should be smooth in comparison to that of the gas and young stars. Using this feedback prescription \cite{oklopcic2017} perform a case study of giant clumps in a massive, $z=2$ disk galaxy and, similarly, find no evidence in their simulations for net inward migration of clumps and predict by $z=1$ a smooth, stellar morphology. Using the NIHAO simulation, \cite{buck2017} find a similar lack of long-lived clumps and suggest that clumps in high-z galaxies are merely the consequence of variable extinction. The similarity between these simulations is that both incorporate early radiative feedback with high photon trapping coefficients in the dust \citep{hopkins2012,stinson2013}.

Because this gas-rich, turbulent mode of star formation occurs primarily in distant galaxies the effects of redshift must be considered. Distance affects observations in three ways: (1) generating native limits to resolution with current instrumentation, (2) creating practical limits to sensitivity, and (3) shifting spectral features to longer wavelengths, making them impractical to observe with current and near-future instruments. The first two effects, resolution and sensitivity, are assuaged by lensed galaxies \citep{jones2010,livermore2012,livermore2015,cava2018,dz2017}. However, in the vast majority of lensed systems the magnification occurs in only one direction, which complicates structural analysis. Furthermore, observations of lensed galaxies can still be impacted by redshift-related effects. For example, resolved measurements of galaxies at $z>2$ cannot be undertaken at $\lambda > 500$nm using current facilities, which significantly challenges efforts to measure the stellar masses of clumps, even in lensed systems. 

The optimal wavelength for constraining the masses of clumps is the near infrared (IR). Mass-to-light ratios estimated from near-IR observations are less sensitive to degeneracies in extinction, age and metallicity than are rest-frame visible wavelength observations \citep{bdj2001}. High-resolution imaging of very clumpy star-forming galaxies at low redshifts would provide a robust (and highly complementary) viewpoint on the physics of the high-redshift population. \cite{green2014} presented the DYnamics of Newly-Assembled Massive Objects (DYNAMO) Survey, which is comprised of 95 nearby ($z\sim 0.06-0.08$ \& $0.12-0.16$) galaxies which closely resemble high-$z$ clumpy systems in terms of their kinematic and star formation properties. The DYNAMO sample has been the subject of a number of follow-on investigations (for example, \citealt{green2014,bassett2014,bassett2017,fisher2014,fisher2017a,fisher2017b,white2017,oliva2018,fisher2019}) which test the similarity of these local galaxies to their potential high-redshift counterparts. The most notable exploration of this theme is presented in \cite{fisher2017a}, who used high-resolution H$\alpha$ maps (from \textit{Hubble Space Telescope}; hereafter, \textit{HST}) to confirm that DYNAMO galaxies exhibit the same clumpy morphology observed at high-redshift. As is the case with massive galaxies at high-redshift, a large fraction of DYNAMO systems appear kinematically to be turbulent disks ($\sigma_{gas} \sim$20-100 km/s; \citealt{green2014,bassett2014,bekiaris2016}) with high molecular gas fractions ($f_{gas} \sim$0.2 - 0.6; \citealt{fisher2014,fisher2019,white2017}). 

In this paper, we build on the work of \cite{fisher2017b} by investigating the existence and properties of stellar clumps in G04-1, a galaxy from DYNAMO, using adaptive optics (AO)-enabled NIRC2 $K_{P}$-band observations\footnote{In this manuscript, the term ``$K$-band" refers to observations taken with the NIRC2 $K_{P}$ filter ($\rm \Delta\lambda = 1.948 - 2.299 \mu m$).} Due to its wealth of previous observations and its clear classification as a highly star forming, clumpy, turbulent, gas-rich disk, G04-1 makes an ideal candidate for probing the nature of stellar clumps. 

This paper is structured as follows: in \S\ref{sec:2}, we provide an overview of these targets and describe the new NIRC2 and \textit{HST} observations. In \S\ref{sec:3}, we describe the methods we have used for identifying stellar clumps and calculating clump properties (such as mass and color) in our imaging data. In \S\ref{sec:4}, we present our results and discuss them in context. \S\ref{sec:5} summarizes our  findings. 

Throughout this paper, we assume a cosmology where $H_{0}$ = 67 km s$^{-1}$ Mpc$^{-1}$, $\Omega_{M} = 0.31$, and $\Omega_{\Lambda} = 0.69$.


\section{Observations}\label{sec:2}

\subsection{Known properties of G04-1}
Galaxy G04-1 is a member of the greater DYNAMO sample \citep[originally presented in][hereafter referred to as DYNAMO-I]{green2014}. DYNAMO is an H$\alpha$ IFU survey of local ($z\sim 0.07$ and $z\sim 0.12$) galaxies which have been selected from the Sloan Digital Sky Survey (SDSS, \citealt{york2000,blanton2017}) to be H$\alpha$-luminous (in the top 1\% of H$\alpha$ emitters in the local universe, based on fiber luminosity; $\overline{SFR}\sim$ 11 $\Msun$ yr$^{-1}$). Located at $z\sim 0.1298$, G04-1 has an integrated stellar mass of 6.47$\times 10^{10}$ $\Msun$ and H$\alpha$-derived SFR of about 15 $\Msun$yr$^{-1}$ (DYNAMO-I). As is the case for G04-1, a large fraction ($\sim$84\%) of DYNAMO galaxies appear disk-like and about half are located on the Tully-Fisher relation \citep{green2014}. \cite{fisher2017a} produce a surface brightness profile from a high-resolution \textit{HST} continuum map for G04-1 and find that the system is well-fit by an exponential disk + bulge model. Based on the data presented in DYNAMO-I and on follow-up kinematic fitting by \cite{bekiaris2016}, G04-1 is classified as a turbulent rotating disk ($V_{\rm circ}=264 \pm 6 \kms$ and $\sigma_{gas} = 34 \kms$). Recently, \cite{oliva2018} have published high-resolution (100 - 400 pc scale) AO-assisted kinematic maps of DYNAMO galaxies imaged in P$\alpha$ with Keck-OSIRIS. For G04-1, these authors find an integrated velocity dispersion of $\sigma \sim 51 \kms$. In addition, these authors observe multiple high-dispersion ($\rm \sigma_{max}\sim 71 \kms$) peaks in the P$\alpha$ map. G04-1 is also observed to be gas-rich; CO line fluxes reported by \cite{fisher2014} (using the Plateau de Bure Interferometer observations of the CO[1-0] transition) estimate a baryonic gas mass fraction of $\sim$31\% and a depletion time of about 1.9 Gyr \citep{white2017}. Like many of the systems commonly observed at high-$z$, G04-1 is morphologically clumpy. Using 100pc resolution \textit{HST} $H\alpha$ maps, \cite{fisher2017a} identify 13 massive star forming clump regions within the galaxy.

\subsection{NIRC2 K-band imaging}
G04-1 was observed with the NIRC2 imaging camera on Keck II using \textsc{widecam4} ($\sim$0.04"/pix scale) with laser guide star adaptive optics correction and with the $K_{P}$ filter ($\rm \lambda_{c} = 2.124\ \mu m$). Observations took place on 2016-21OCT as part of the 2016B observing cycle (program W140N2L) for 1.75 hours. 

Observations were reduced using fairly standard methods. First, all raw frames were corrected for known bad pixels using the Keck-NIRC2 bad pixel map. Dark current was removed from science and flat frames by subtracting off a scaled master dark frame. Both lamp-on and lamp-off flat frames (8 each) were recorded on the night of observation, and these frames were median-combined separately (with sigma-clipping) and then subtracted (on-off) to construct a flat field frame. This was divided through all science frames to flat-correct the data.

Keck/NIRC2 images show interference fringes, which are easily detected in flat-corrected frames. We modeled this pattern by first selecting all co-temporal science frames within which all sources were masked. In most cases this corresponded with the three frames associated with a single dither pattern. These masked frames were median-combined and scaled (to an appropriate background level) to produce frames which were subtracted off the individual images to perform removal of sky and fringes in a single step. The reduced science frames were individually corrected for distortion effects using IRAF's \textsc{drizzle} task in combination with  X \& Y \textsc{wide} camera distortion solutions provided by H. Fu.\footnote{https://www2.keck.hawaii.edu/inst/nirc2/dewarp.html} Image registration and sub-pixel offsets between the dithered frames were calculated with IRAF's \textsc{xregister} task. Finally, these offset values were used with \textsc{drizzle} once again to produce a final co-added image. 

Our G04-1 frames contained a ghost image of the Keck primary mirror positioned in the lower-right quadrant of the array. In some cases ($<30\%$ of frames) this image (which appeared static over the course of the night's observations) overlapped partially with the dithered position of the galaxy. Where possible (i.e. where the source is in a different quadrant) the region with this feature was left unmasked so as to remove it during the sky subtraction process described above.

\begin{figure*}[htb!]
\begin{center}
\includegraphics[scale=0.59]{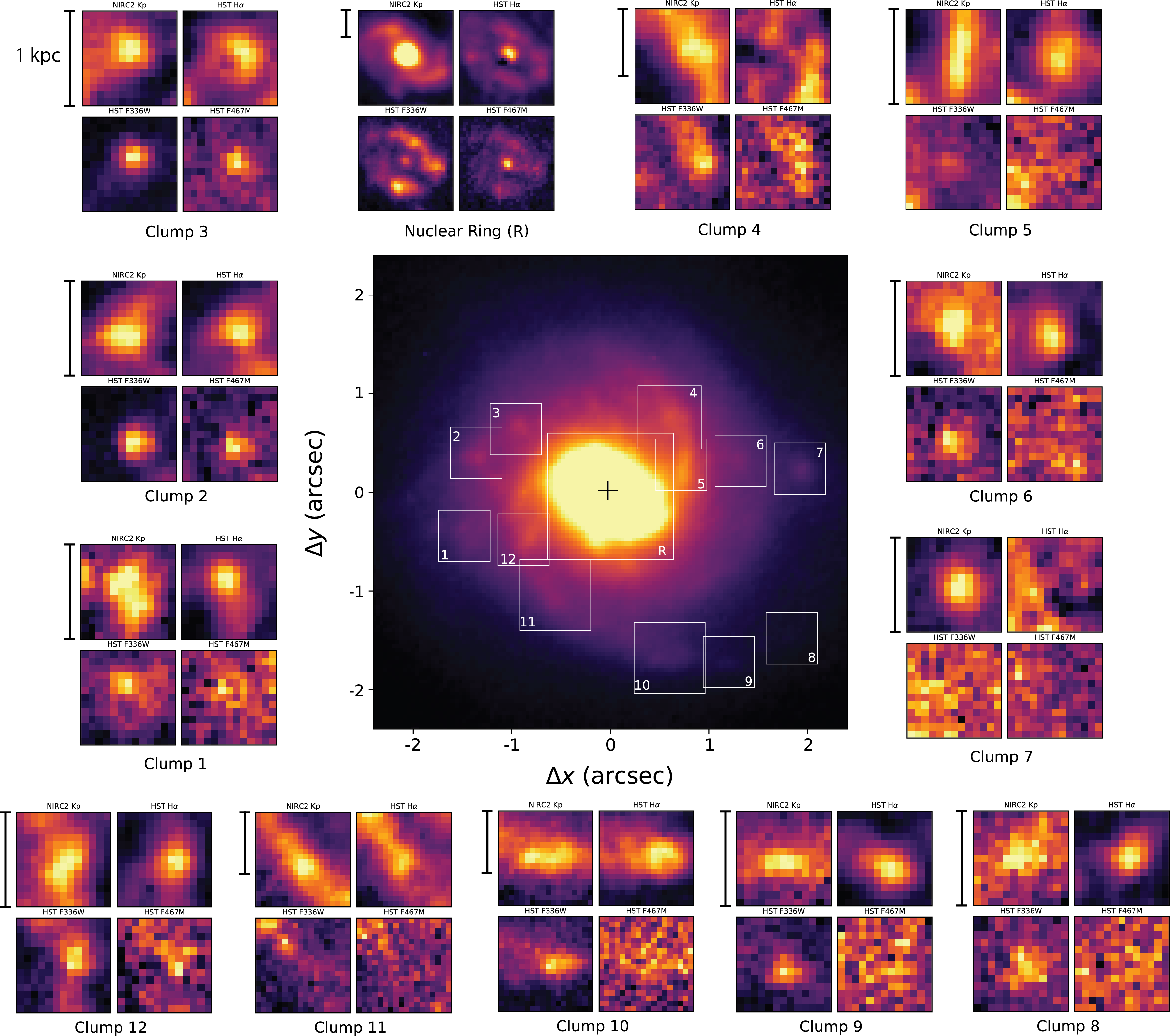}
\end{center}
\caption{Multi-band Imaging of Stellar Clumps in G04-1. In the center panel, we show the reduced, co-added (n=21) image of G04-1 from Keck NIRC2 observations in K$_{\rm P}$-band. White boxes within the center panel denote individual clump regions with labels corresponding to the Clump ID numbers provided in Tables \ref{tab:phot} \& \ref{tab:prop}. Surrounding this center panel, we include enlarged maps of each clump region in K-band (with the disk light removed) and in our \textit{HST} $H\alpha$ (from \citealt{fisher2017a}), F336W, and F467M data sets. Due to the complex structure near the center of G04-1, clumps 13, 14, \& 15 are shown in a single region labeled the "nuclear ring". The on-sky region depicted by white boxes in the center panel match that of each of the corresponding clump's enlarged panels. For reference: cutouts for clumps 1, 2, 3, 5, 6, 7, 8, 9, \& 12 are represent on-sky sizes of $\sim 1$ kpc across. For clump 4, this cutout represents a size of $\sim 1.4$ kpc and for clumps 10 \& 11, about $\sim 1.5$ kpc. Finally, the width of the white box enclosing the nuclear ring (R) region is $3.7$ kpc in size.}
\label{fig:clump_apertures}
\end{figure*}

\subsection{HST F336W and F467M Observations}\label{sec:HST}
We obtained observations (Proposal ID \#15069) of G04-1 using the Wide Field Camera on the Advanced Camera for Surveys (WFC/ACS) on the \textit{Hubble Space Telescope}. Observations were performed using the F336W ($U$) wide and F467M (Stromgen $b$) medium filters on 2018-02-26 for a total integration time of 0.5 and 0.2 hrs, respectively. For further description of the broad-band \textit{HST} images we refer the reader to \cite{lenkic2021}. Charge transfer effects in WFC images were mitigated both by placing the target close to the readout edge of the image, and also by using a post-flash to raise the image background to 12 e-/pix. All images were reduced using the standard \textit{HST} pipeline and combined using \textsc{drizzle}.

\section{Properties of Stellar Clumps in G04-1}\label{sec:3}

\subsection{Identification of Stellar Clumps}

Clumps were identified in our galaxy maps as follows. First, we constructed a model of our AO point-spread function (PSF) by combining all imaging of all point sources from a single observing night (described in more detail in \S3.3). We then convolved the co-added K-band image with a Gaussian kernel with a width of 10x the full-width-at-half-maximum (FWHM) of the PSF (determined by a double Moffat function fit; details on this derivation are provided in \S 3.3). This corresponds to a kernel size of $\sim$1.2 arcseconds. This convolved galaxy image was then divided through the un-convolved image to produce a ratio map from which we identified stellar clumps.

Next, all ratio map pixels above a given threshold (defined as 4x the standard deviation of the the intraclump regions) were identified as peak-value pixels. We then imposed the constraint that all peak-value pixels represented local peaks in the flux map. Here, we defined the local regions surrounding the peak values as boxes 10$\times$10 pixels ($\sim$1 kpc $\times$ 1 kpc at $z\sim$0.1; motivated by the typical clump sizes observed in G04-1 by \citealt{fisher2017b}) in size. If higher-value pixels were found within in this local region, the location of the flux peak was shifted accordingly. It is emphasized that this technique assumes a minimum separation between candidate clumps (the box size). Once local peaks were identified, we determined the center of the clump candidates by centroid fitting a 2D Gaussian + constant to a cutout region of the map surrounding the local peak pixel. We then imposed a final constraint that all clump candidates were least as large in size as the FWHM of our combined AO PSF (about 3$\times$3 pixels in area). Within G04-1, this process identified the 15 stellar clumps. These clumps are shown in the center panel of Fig \ref{fig:clump_apertures}, where individual clump regions are indicated by white boxes. In this figure, we also include maps of each of the identified clumps in the \textit{HST} H$\alpha$ (from \citealt{fisher2017a}), F336W, and F467M data sets.

The number of detected clumps is inherently connected to the detection threshold. Decreasing the detection threshold of clumps to 3-$\sigma$ would increase the number of potential clumps in G04-1 to $\sim$30 from 15. If we make a more strict definition of clumps using a threshold of 5-$\sigma$ this decreases the number of clumps to $\sim$10. However, we observe that our algorithm fails to identify fainter and smaller clumps (such as ID=1 and 9) which are otherwise easily identified by eye. Of course, we acknowledge that this illustrates there exists a systematic uncertainty in this choice.

We recognize that the clumps we report here in G04-1 do not likely represent a full statistical sample of all star clusters in the galaxy. G04-1 is a galaxy with diverse forms of structure, including multiple clumps (as identified in this paper), spiral arms, and a bright nuclear ring. Our aim in this paper is simply to determine if a rotating disk galaxy with large, observed clumps of star formation likewise has larger structures (i.e. “clumps”) in maps of starlight, and (if so) where are these masses located and to report their corresponding masses. We acknowledge that constructing a catalog of all observable structures within the galaxy, spanning a wide range of spatial scales, is beyond the scope of this work.

In summary, in the present paper we define stellar ``clumps" to be regions within the galaxy disk which (1) represent local flux peaks (2) are entirely comprised of pixels above a given threshold level (4x the background of the disk) and (3) are at least 3$\times$3 pixels in size (the FWHM of the AO PSF). A list of the photometric properties of all regions meeting these criteria for G04-1 is given in Table \ref{tab:phot}.

\begin{figure*}[htb!]
\begin{center}
\includegraphics[scale=0.66]{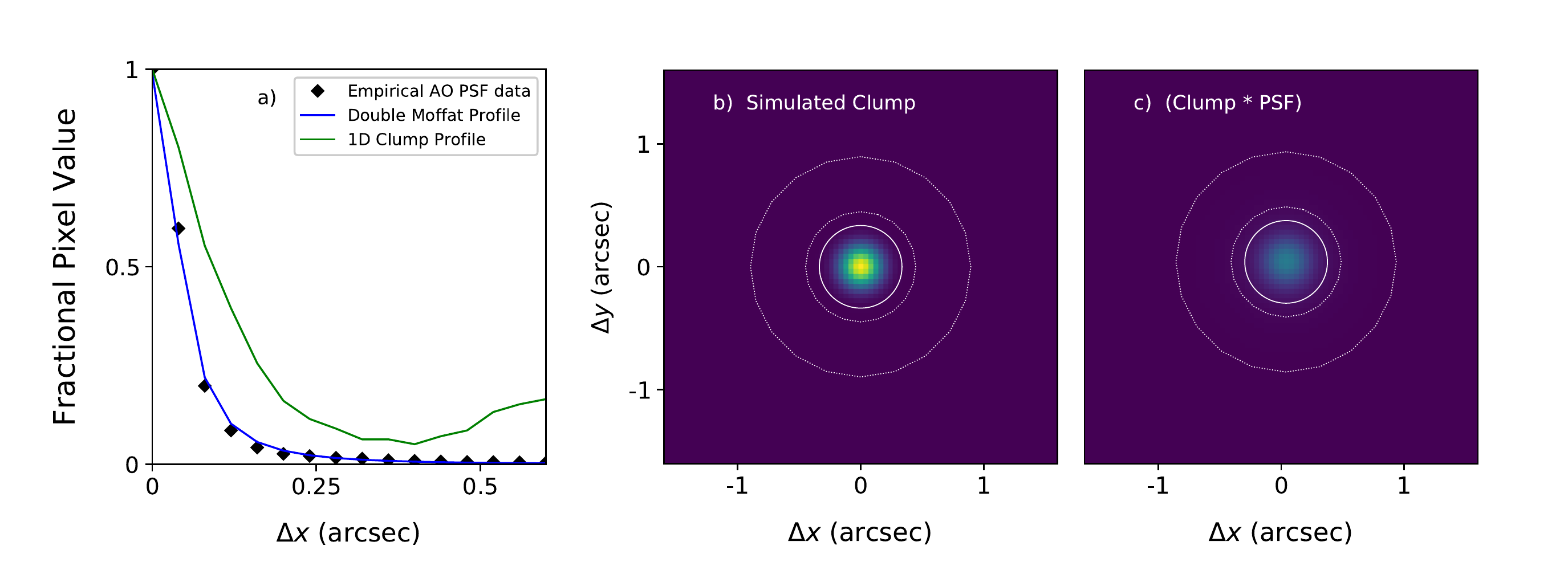}
\end{center}
\caption{Exploring AO PSF effects on clump photometry. In panel 2a  the disk-subtracted 1D brightness profile of the combined AO point-spread function (black diamonds; see \S 3.3) is shown in comparison with the brightness profile for a typical clump (in green; for Clump ID = 2) and the double Moffat best-fit to the data (in blue). Panels 2b \& 2c illustrate the effects of convolving the observed point-spread function with a simulated stellar clump. Dashed white circles show the regions defining the photometric aperture and background annulus. These regions have been used to determine the fraction of clump light lost due to the broad wing component of the AO PSF when performing photometry ($\sim$30\%).}
\label{fig:AOPSF}
\end{figure*}

\subsection{Determination of Clump Fluxes}
We calculated the flux of clumps in two ways. First, we simply integrated the flux of all pixels within a defined aperture centered on each clump (apertures were placed at the locations determined in the previous section). In most cases (12 of 15), the clumps identified in our K-band map could be directly associated with clumps observed in H$\alpha$ by \cite{fisher2017a}. In these cases, we utilized the apertures described in that work in our flux calculations. In four cases (Clump IDs: 4, 7, 11, \& 15) we identified NIR clumps with no obvious H$\alpha$ counterpart. For these clumps, appropriate apertures were estimated using the observed sizes in the ratio map. 

We note that while Clump ID = 7 is well-detected in K-band imaging it is unique in that it has no observable counterpart (i.e. a feature with similar morphology and position) in any of the \textit{HST} data sets presented in this paper. Suspecting that this object is not actually a clump in G04-1 but the serendipitous detection of a background infrared source, we have performed a sky coordinate search using the online search tool \textsc{Vizier} \citep{vizier} but were unable to find a known object for which we could directly associate this emission. Unable rule out that this feature is indeed a stellar clump (without significant $\rm H\alpha$ emission) in G04-1, we include it in the analysis and discussion sections which follow.

Next, a local disk background subtraction was performed to remove light contributed from the diffuse disk component. The local disk background value was determined from a region surrounding each clump (an annular aperture of area equal to that of the clump aperture). Nearby clump pixels that fell within this annular region were omitted from the background estimate. This local background component (defined as the mean background value multiplied by the area of the clump aperture) was then subtracted off the flux of the clump, $F_{K_{P}}$, to obtain a disk-subtracted flux estimate ($F_{K_{P},diskcorr}$). These values are listed in Table \ref{tab:phot}. While disk subtraction is typically performed via bulge-disk decomposition methods, attempts (using both \textsc{GalFit} and \textsc{ProFit} softwares; \citealt{galfit,profit}) to fit a bulge and disk component to the K-band image of G04-1 left substantial structure within the galaxy (i.e. the ring, spiral arms, and bright peaks in flux associated with clumps), which were difficult to model with tolerable residuals. Therefore, we chose to adopt a more clump-centric approach to the removal of the background component. 

\begin{figure*}[htb!]
\begin{center}
\includegraphics[scale=0.47]{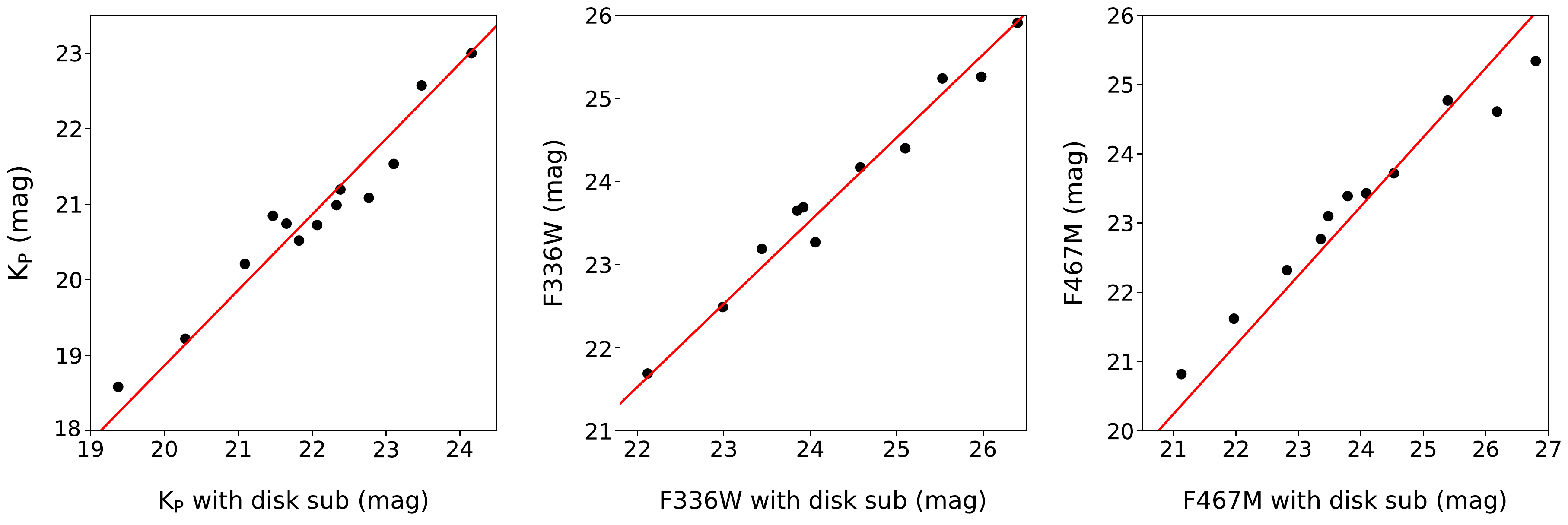}
\end{center}
\caption{An assessment of \textit{Keck} NIRC2 $K_{P}$-band and \textit{HST} F336W and F467M filter AB magnitudes (with and without local disk subtraction) for clumps in G04-1 (unity slope reference lines shown in red). The near 1-to-1 relationship between disk corrected and non-disk corrected magnitudes indicates effects of correction for disk light are independent of clump position in the galaxy.}
\label{fig:Kp_corr}
\end{figure*}

G04-1's guide star was used to derive a zero-point value for converting counts/s to flux. All images of the guide star were reduced using the methods described above, aligned by centroiding, and combined to produce a final image. The star (2MASS J04122098-0555104) has an entry in the Two Micron All-Sky Survey Point-source Catalog (hereafter 2MASS; \citealt{skrutskie2006}). We derived a $K_{P}$ magnitude for the guide star using its $K_{S}$ magnitude (from 2MASS) in combination with the flux ratio of the bandwidths between the $K_{P}$ and $K_{S}$ filters. This, in combination with our imaging of the star, was used to produce a conversion value between counts/s and flux.

\subsection{AO-PSF Effects on Photometry}
As described in $\S$2.2, our observations of G04-1 incorporate laser guide star AO correction. The shape of the AO PSF is known to vary in both time and with observing conditions and is generally modeled as a near diffraction-limited core with a broad, seeing-limited halo/wing component. Except in cases with very high Strehl ratios, a significant fraction of the light is shifted into the wings. The impact of the multi-component PSF of AO systems may therefore act to systematically reduce the measured flux of clumps. Assuming standard performance of the Keck AO system, we expect a Strehl ratio of 0.25 in the $K_{P}$ filter, based on our tip-tilt star magnitude (R$\sim$17 mag and 2MASS $K_{S}=15.7$ mag) and distance to our target ($r_{\rm offset} \sim 28.4\arcsec$). The Strehl ratio associated with this set of observations was independently estimated by comparing imaging of point sources observed the same night with a model of the diffraction-limited PSF. We find Strehl ratios are very similar (0.26, on average) to that supplied by Keck and therefore a value of 0.25 is used in all of our flux-magnitude calculations. Here, we explore some of the recovery biases associated with performing photometry on our AO-enabled $K_{P}$-band maps.

In order to increase the signal-to-noise of the faint component, we have constructed a model of the AO PSF through a median combination of all of the point sources observed (scaled by exposure time) on 2016-21OCT. These 21 frames correspond to a combined integration time of 1.4 min. In Fig. \ref{fig:AOPSF}a, we provide a plot of the azimuthally-averaged 1D surface brightness profile of the combined source (shown as black diamonds). We find that this AO PSF model from our observations is well fit by a double Moffat function with best-fit parameters (amplitude, core width, power index) of (0.91, 0.06$\arcsec$, 1.56) and (0.08, 0.13$\arcsec$, 1.21) for the core and halo components, respectively (shown as a blue solid line). Note: the centering of our double Moffat profile was fixed to the source origin. For comparison, we also include in Fig. \ref{fig:AOPSF}a a scaled 1D profile of a characteristic clump (ID: 2) in G04-1. The FWHM corresponding to this double Moffat profile is $\sim 0.1 \arcsec$, which is consistent with that expected in K-band at this observing site and given the Strehl ratio assessed above. 

Due to image construction effects (i.e. the \textsc{Drizzle} algorithm), the empirical PSF of the final, co-added \textit{Keck}-NIRC2 image is likely slightly underestimated by this point-source FWHM value. Although, we expect the difference between these two resolution metrics to be small ($<10\%$). One method of inferring the spatial resolution of this image would be to assess features observed within the galaxy which appear smaller than clumps. The dominance of star light in the disk, however, makes the search for small structures in the galaxy prohibitive. Nevertheless, in the field of the final image of G04-1 we do observe one small source for which we estimate a FWHM of $\sim0.15$ arcsec. This provides additional confidence in the FWHM estimate we utilize here to explore the effects on flux loss on our K-band clump photometry. We note that our \textit{HST} F336W ($U$) and F467M ($b$) data sets are also processed by \textsc{Drizzle} and PSF modeling using \textsc{TinyTim} predicts FWHM values in this bands of $\sim0.1$ arcsec.\footnote{\textsc{TinyTim} is a PSF modeling software package developed by J. Krist and R. Hook. Access to this software is available through the Space telescope Science Institute (STSCI)  HST instrumentation website.} Within our final F336W image, we observe a number of point sources in the field surrounding the galaxy. From averaging the on-sky sizes of a few of these sources, we estimate a spatial resolution for our \textit{HST} data of $\sim 0.09$ arcsec, which is slightly better in comparison to our K-band maps.

We are interested primarily in estimating the fraction of the light originating from stellar clumps that is shifted beyond our flux apertures (i.e. the ``light loss") due to the broad wing component of the AO PSF. We perform a simple simulation of the effects of the PSF on a stellar clump, which we show in Fig. \ref{fig:AOPSF}. Here, we model our clump with a 2D Gaussian function where the FWHM is set by an average, circularized clump radius ($\sim0.13\arcsec$ or $\sim$300 pc) from our sample in Table \ref{tab:phot} (Fig. \ref{fig:AOPSF}b). We note that many of the clump sizes are consistent with sizes from \textit{HST} H$\alpha$ imaging, and do not suffer from broad components of the PSF. We then convolved this 2D clump with the previously described double Moffat profile to mimic the smearing effect of our empirical PSF (see Fig. \ref{fig:AOPSF}c). To assess the fraction of light lost, we use a fixed aperture size of $r=3\sigma$ (where $\sigma$ is defined by the double Moffat FWHM) and annulus radii of $r_{in}=4\sigma$ and $r_{out}=8\sigma$ (see Fig. \ref{fig:AOPSF}b \& c) for performing flux and background photometry.

After convolution with our AO PSF model, approximately 66\% of the clump light remains within the 3$\sigma$ flux aperture, corresponding to a flux/mass correction factor of 1.33. However, in reality these clumps are embedded within the galaxy disk and the broad PSF wings also feed stellar light from the surrounding disk back into our clump apertures. This is likely to reduce flux lost from clumps. Since we have not included an underlying disk component in our simulation, this correction factor of 1.33 represents an upper limit for the fraction of light lost due to PSF effects.

In our simulation, we find that the per-pixel flux contribution to the local background estimate (determined within the aforementioned annulus) is of order $\sim$1\% of the amplitude (i.e. peak pixel value) of the clump. Again, we present this as a rough estimate; clump clustering and bright structures in G04-1 requires estimations of the local disk background to be determined from a range of annuli radii. Moreover, the morphology of stellar clumps is observed in our maps to deviate from simple Gaussians. Nevertheless, we note that clump peak pixel values range between 2-5$\times$ higher than their corresponding local disk values, suggesting that this light represents a minor contribution to our background estimates. 

The authors highlight that the above consideration of the impact of AO-correction on photometry measurements of clumps within G04-1 is really only possible due to the galaxy's location at relatively low redshift. The scale length of star forming clumps in G04-1 is small when compared with the scale length of the disk. This allows us to more cleanly separate out the clumps from the underlying disk of the galaxy. This underscores another unique advantage of studying processes at high-$z$ via targets in the DYNAMO sample.

\subsection{Clump Stellar Masses}
It is commonly assumed that stellar mass density varies in lock-step with near-infrared surface flux density, but we have used our in-hand visible-wavelength \textit{HST} photometry to try to refine our mass estimates using population synthesis models. Integrated and clump-scale mass-to-light ratios for G04-1 were estimated using the stellar population synthesis code \textsc{galaxev}, which comprises a library of evolutionary models computed using the isochrone synthesis codes of \cite{bc2003}.  We used spectral energy distribution (SED) models from the \cite{bc2003} code base (incorporating their 2011 update), assuming a Chabrier initial mass function \citep{chabrier2003}. As our goal was to derive mass-to-light ratios for isolated clump regions, we modeled the star formation history of clumps as a simple stellar population (i.e. a delta-burst). Galaxies in DYNAMO (including G04-1) are observed to have metallicities which are slightly sub-solar (determined from the [NII]/$H\alpha$ ratio \citep{pp2004} from SDSS spectra) and an BC2003 SED model consistent with this was chosen from the BaSeL 3.1 spectral library \citep{bc2003}.

Clump colors vary to a small degree across the galaxy disk, and we have used the \textit{HST} data described in \S\ref{sec:HST} to construct ($U-b$) maps (using F336W and F467M). Using the apertures defined in calculating $K_{P}$-band magnitudes on these maps, we calculated a ($U-b$) color index (with and without disk-subtracted magnitudes) for each of the clumps in G04-1. The standard deviation of the clump colors was 0.36 (in AB mag units). Using the BC03 software, we then modeled the evolution of the $K_{P}$ mass-to-light ratio as a function of observed ($U-b$) color to derive clump-specific mass-to-light ratios. The stellar mass of each clump is then defined as simply the product of the NIR flux and the BC03-derived mass-to-light ratio of the clump. Derived ($U-b$) clump colors, mass-to-light ratios, and masses are all listed in Tables \ref{tab:phot} \& \ref{tab:prop}.

The simple stellar population's observed magnitudes in our selected filters are evaluated at equally spaced time steps ($\Delta log(\rm Age)=$ 0.05 yr, beginning at $10^{5}$ yr) and thus we are able to track the color evolution of the burst with time. This provides us a reasonable metric for estimating the ages of the clumps, using simply their individual $U-b$ colors, which we include on the right-hand axis of Fig. \ref{fig:mass_vs_dist} (top panel). We note that a particular age of a clump should be interpreted carefully in terms of the clump lifetime. \citet{bournaud2014} outlines the difficulties associated with accurately measuring the lifetime of a clump based solely on photometric ages. Clumps can have complex star formation histories and their simulations show that if clumps are long lived, they will continually rejuvenate (i.e. experience subsequent bursts) and young stars dominate the age measurement. However, \citet{bassett2014} reports (based on absorption line analysis) an age range of $60-500$ Myr for the entire galaxy of G04-1 which is consistent with the clump ages we derive here using stellar population synthesis modeling.

\begin{figure}[htb!]
\begin{center}
\includegraphics[scale=0.56]{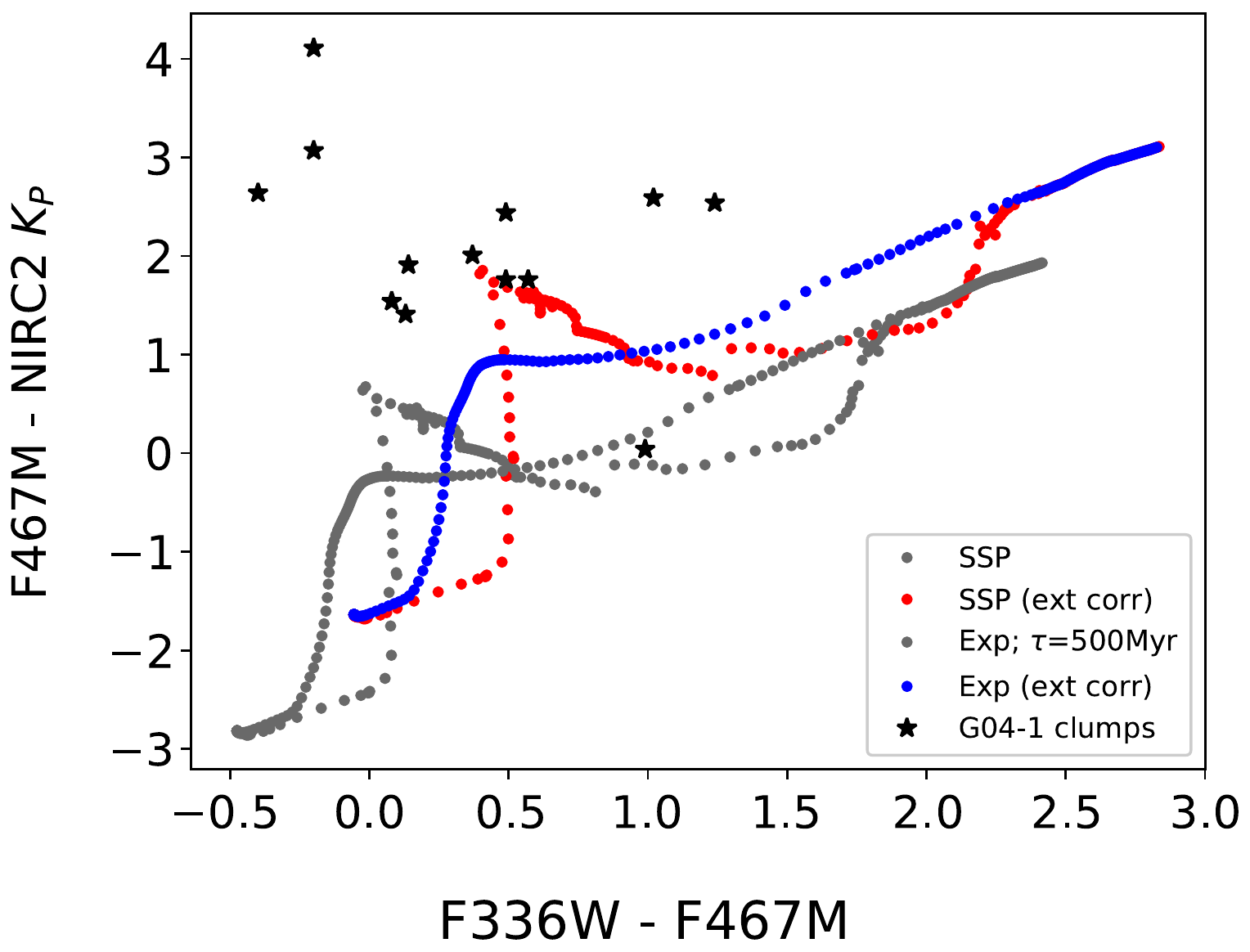}
\end{center}
\caption{In this figure, the observed $(U-b; F336W - F467M)$ and $(b-K)$ colors of clumps in G04-1 are compared to those of two example synthetic stellar populations (generated using \textsc{galaxev} and stellar models from \citealt{bc2003}). Here, the positions of stellar clumps are shown as black stars. The evolutionary tracks of two star formation histories, a simple delta-burst (SSP) and an exponentially-declining SFR (Exp; with an e-folding timescale of $t=500 Myr$) are provided for reference. SSP and Exp tracks in red and blue, respectively, include colors with extinction applied, assuming a \cite{calzetti2000} extinction law. Their counterparts plotted as grey do not account for effects from extinction.}
\label{fig:clumpcolors}
\end{figure}

As described above, we derive mass-to-light ratios for clumps via a method which assumes that 1) the observed colors reflect that of a single burst of star formation and 2) that the clumps extinction values are similar to that of the disk (e.g. in using disk-subtracted values). Note: the assumption that clump extinction is similar to the disk is supported by observations of the $H\alpha$-to-$P\alpha$ flux ratio in G04-1, which is found to be roughly constant (Bassett et al 2017). Given the observed stability of the K-band mass-to-light ratio, the uncertainties associated with estimating stellar masses via ($U - b$) color are not expected to be significant (e.g. a small variation in clump color is observed across the galaxy). However, as magnitude information is available for clumps in three bands ($U$, $b$, and $K_{P}$) we can compare the observed ($U-b$) and ($b-K$) colors with those generated from stellar population synthesis modeling. This comparison with two, distinct star formation histories (a ``delta-burst" SSP and an exponentially-declining SFR) is shown in Fig \ref{fig:clumpcolors}. In this comparison, we utilize colors calculated from disk-subtracted magnitudes to minimize K-band reddening from the underlying stellar disk in which the clumps are embedded. We incorporate the effects of extinction and reddening via the \cite{calzetti2000} dust extinction law assuming $A_{H\alpha}\sim1-1.6$ (derived for G04-1 by \citealt{bassett2017}) and $A_{V}\sim 1$. Tracks which account for extinction are shown for the SSP and Exponential star formation histories in red and blue, respectively. We find that four clumps exhibit both ($U-b$) and ($b-K$) colors which map well to those generated by BC03 SSP modeling. Five clumps are within a magnitude in color from either the SSP or Exp BC03 tracks. Three clumps are vertically offset by $>$1 mag from both tracks, appearing much more red in ($b-K$) color than predicted by BC03.

These results are challenging to interpret as it remains unclear how well-represented these clump populations are by the simple star formation histories presented here. Indeed, the star formation histories of clumps in high-$z$ turbulent disk galaxies are very poorly constrained. Results from simulation work (e.g. \citealt{bournaud2014}) suggest that some clumps may undergo multiple bursts of star formation over the course of their lifetime. While actively star forming clumps are dominated by young stars, they may be holding onto a redder, more evolved stellar population which might explain them appearing significantly more red. We note, however, that attempts to fine-tune our models for different star formation histories (a SSP and exponential SFR with timescales of $100< \tau < 500$ Myr) result in an overall range in the average K-band mass-to-light ratio of 0.1 - 0.15. Further efforts to fine-tune are likely beyond the scope of this paper. Nonetheless, this shows that our uncertainty in model results in a small addition to the overall systematic uncertainty associated with stellar mass. This underscores the clear advantage of estimating clump stellar masses using K-band photometry.

\section{Results \& Discussion}\label{sec:4}
\begin{figure}[htb!]
\begin{center}
\includegraphics[scale=0.56]{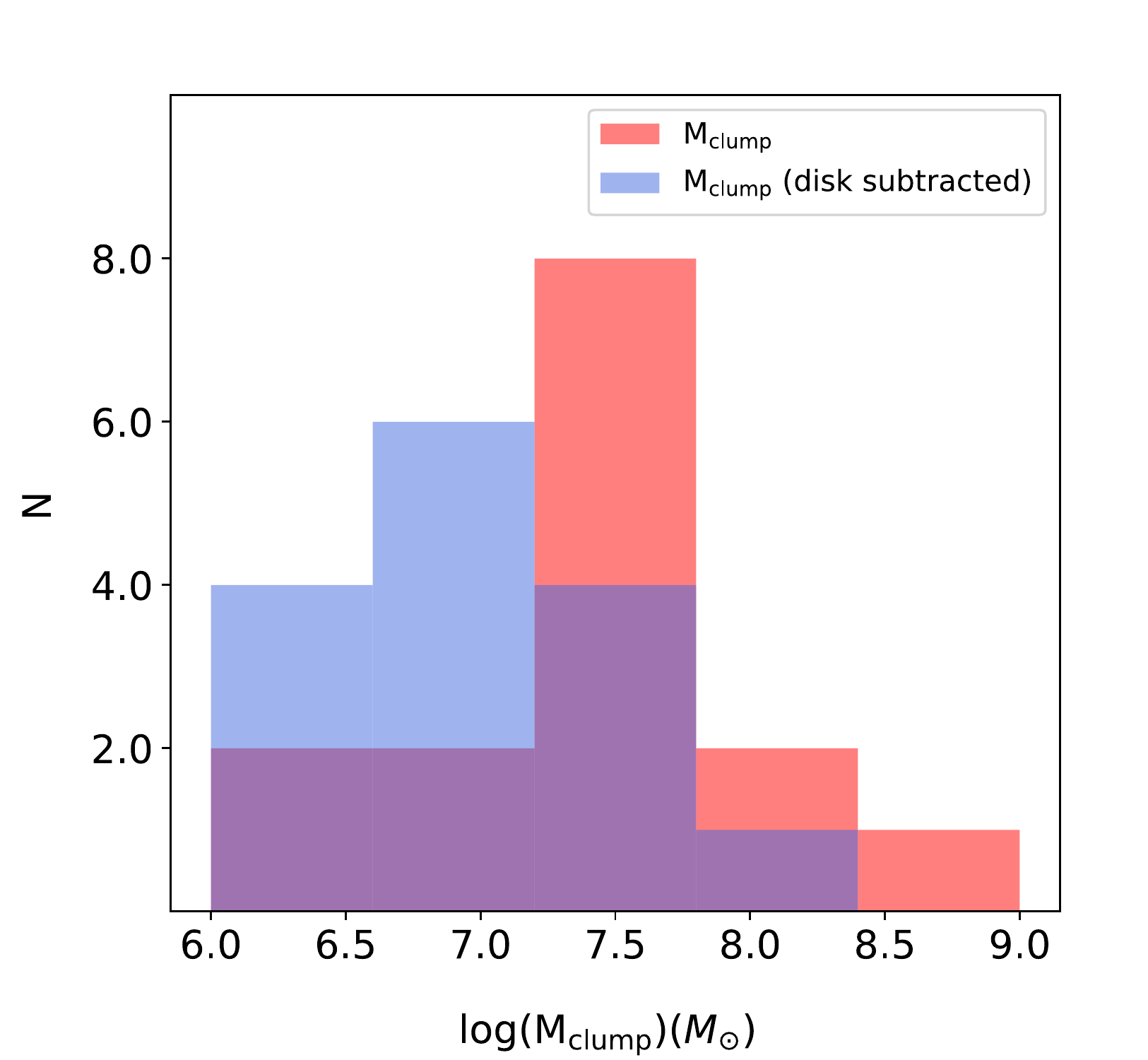}
\end{center}
\caption{The distribution of clump stellar masses in G04-1. The red and blue histograms represent stellar masses determined for the clumps described in Table \ref{tab:phot} both with and without local disk background subtraction.}
\label{fig:Mstar_hist}
\end{figure}

Figure \ref{fig:Mstar_hist} shows a histogram of the full range of clump masses, both before and after background flux subtraction is performed. On average, subtraction of the local disk component reduces clump masses by around 50\%. The average stellar mass of the clumps before (after) background subtraction for the 15 identified clumps in G04-1 is $5.69\pm 1.8\times 10^{7}\Msun$ ($2.06\pm 0.7\times 10^{7}\Msun$). The highest and lowest mass clumps (Clump IDs 14 \& 9) correspond to disk-subtracted masses of 27.0 and 0.36 $\times 10^{7}\Msun$, respectively. If we incorporate the correction for light-loss due to wings of the AO PSF (see \S 3.3), the maximum mass for clumps within G04-1 may be as high as 3.6 $\times 10^{8} \Msun$. We note that the fractional drop in mass from the background subtraction is quite consistent across clumps (i.e. uncorrelated with clump brightness or position in the disk), as shown in Fig. \ref{fig:Kp_corr} where we directly assess the effect of disk-subtraction on calculated magnitudes for clumps in the $K_{P}$, F336W ($U$), and F467M ($b$) datasets.

\begin{deluxetable*}{ccccccccccc}
\tabletypesize{\scriptsize}
\tablecaption{G04-1 Clump Photometry\label{tab:phot}}
\tablewidth{0pt}
\tablehead{
\\
\colhead{Clump ID} &
\colhead{$K_{P}^{\dagger}$} &
\colhead{$K_{P, diskcorr}^{\dagger}$} &
\colhead{$F_{K_{P}}$} &
\colhead{$F_{K_{P, diskcorr}}$} &
\colhead{$M_{\star}$} &
\colhead{$M_{\star, diskcorr}$}
\\
\colhead{} &
\colhead{(AB mag)} &
\colhead{(AB mag)} &
\colhead{($10^{-14}$\ erg/s/cm$^{2}$)} &
\colhead{($10^{-14}$\ erg/s/cm$^{2}$)} &
\colhead{($10^{7}$\ $\Msun$)} &
\colhead{($10^{7}$\ $\Msun$)}
\\
}
\startdata
1 & 21.08 & 22.77 & 2.01 $\pm$ 0.41 & 0.43 $\pm$ 0.10 & 2.59 $\pm$ 0.79 & 0.55 $\pm$ 0.18\\
2 & 21.19 & 22.38 & 1.81 $\pm$ 0.37 & 0.61 $\pm$ 0.13 & 0.85 $\pm$ 0.50 & 0.28 $\pm$ 0.19\\
3 & 20.52 & 21.82 & 3.38 $\pm$ 0.68 & 1.02 $\pm$ 0.21 & 1.98 $\pm$ 1.07 & 0.60 $\pm$ 0.30\\
4 & 19.70 & 21.26 & 7.22 $\pm$ 1.45 & 1.88 $\pm$ 0.39 & 12.7 $\pm$ 3.31 & 3.31 $\pm$ 0.89\\
5 & 19.52 & 21.14 & 8.47 $\pm$ 1.70 & 1.91 $\pm$ 0.40 & 14.9 $\pm$ 3.88 & 3.37 $\pm$ 0.89\\
6 & 21.53 & 23.11 & 1.33 $\pm$ 0.27 & 0.31 $\pm$ 0.07 & 1.56 $\pm$ 0.50 & 0.37 $\pm$ 0.12\\
7 & 20.72 & 22.07 & 2.80 $\pm$ 0.57 & 0.81 $\pm$ 0.19 & 3.28 $\pm$ 1.06 & 0.95 $\pm$ 0.33\\
8 & 23.00 & 24.16 & 0.34 $\pm$ 0.07 & 0.12 $\pm$ 0.04 & 0.36 $\pm$ 0.13 & 0.13 $\pm$ 0.05\\
9 & 22.57 & 23.48 & 0.51 $\pm$ 0.11 & 0.22 $\pm$ 0.05 & 0.36 $\pm$ 0.17 & 0.16 $\pm$ 0.08\\
10 & 20.85 & 21.47 & 2.50 $\pm$ 0.51 & 1.41 $\pm$ 0.30 & 2.64 $\pm$ 0.91 & 1.49 $\pm$ 0.52\\
11 & 20.74 & 21.65 & 2.75 $\pm$ 0.55 & 1.19 $\pm$ 0.25 & 2.90 $\pm$ 1.00 & 1.26 $\pm$ 0.44\\
12 & 20.99 & 22.33 & 2.19 $\pm$ 0.44 & 0.64 $\pm$ 0.13 & 3.35 $\pm$ 0.93 & 0.97 $\pm$ 0.27\\
13 & 19.22 & 20.28 & 11.2 $\pm$ 2.24 & 4.19 $\pm$ 0.84 & 5.25 $\pm$ 3.45 & 1.97 $\pm$ 1.29\\
14 & 18.58 & 19.38 & 20.1 $\pm$ 4.02 & 9.69 $\pm$ 1.94 & 26.6 $\pm$ 7.95 & 12.9 $\pm$ 3.83\\
15 & 20.21 & 21.09 & 4.50 $\pm$ 0.90 & 2.00 $\pm$ 0.40 & 5.96 $\pm$ 1.78 & 2.65 $\pm$ 0.79\\
\enddata
\tablenotetext{\dagger}{Errors on clump AB magnitudes are approximately 0.24 mag.}
\end{deluxetable*}

\begin{deluxetable*}{ccccccccccc}
\tabletypesize{\scriptsize}
\tablecaption{Observed Properties of Stellar Clumps in G04-1\label{tab:prop}}
\tablewidth{0pt}
\tablehead{
\\
\colhead{Clump ID} &
\colhead{HST ID} &
\colhead{SFR$_{}^{\star}$} &
\colhead{$\sigma^{\star}$} &
\colhead{F336W - F467M$^{\dagger}$} &
\colhead{$K_{P}$ MLR} &
\colhead{$R_{GC}$}
\\
\colhead{} &
\colhead{ } &
\colhead{($\Msun$/yr)} &
\colhead{$\kms$}&
\colhead{(AB mag)} &
\colhead{} &
\colhead{(kpc)}
\\
}
\startdata
1 & 12 & 0.85$\pm$0.15 & 51 & 0.68 & 0.11 & 3.61\\
2 & 13 & 0.62$\pm$0.14 & 51 & 0.30 & 0.04 & 3.35\\
3 & 14 & 0.93$\pm$0.22 & 51 & 0.41 & 0.05 & 2.81\\
4 & - & -  & 51 & 1.15 & 0.15 & 2.47\\
5 & 6 & 1.36$\pm$0.22 & 48.24$\pm$19.8 & 1.15 & 0.15 & 1.90\\
6 & 4 & 0.29$\pm$0.07 & 51 & 0.65 & 0.10 & 3.18\\
7 & - & -  & 51 & 0.65 & 0.10 & 4.60\\
8 & 1 & 0.19$\pm$0.04 & 51 & 0.58 & 0.09 & 5.46\\
9 & 2 & 0.28$\pm$0.05 & 51 & 0.47 & 0.06 & 4.90\\
10 & 3 & 0.43$\pm$0.12 & 51 & 0.55 & 0.09 & 4.06\\
11 & - & -  & 51 & 0.58 & 0.09 & 2.93\\
12 & 10 & 0.39$\pm$0.10 & 51 & 0.74 & 0.13 & 2.36\\
13 & 5 & 1.36$\pm$0.31 & 55.7$\pm$27.62 & 0.30 & 0.04 & 1.05\\
14 & 9 & 2.1$\pm$0.39 & 41$\pm$14 & 0.88 & 0.11 & 0.79\\
15 & - & -  & 51 & 0.87 & 0.11 & 1.16\\
\enddata
\tablenotetext{\dagger}{\textit{HST} F336W-F467M AB color values represent non-disk subtracted colors and have associated errors of on average 0.6 mag.}
\tablenotetext{\star}{Values for clump star formation rates and velocity dispersions have been taken from Fisher et al. (2017b) and Oliva-Altamirano et al. (2018), respectively.}
\end{deluxetable*}

As described in \S1, predictions from simulations for stellar clump masses in turbulent, clumpy disk galaxies depend quite strongly on the detailed feedback prescription assumed by the model. Observations of galaxy clumps (such as those presented here) are therefore useful for testing these models and for providing insight into the dominant forms of feedback. Strong radiative feedback models predict that clumps of gas are disrupted so quickly that the stellar morphology in these galaxies should be relatively smooth \citep{hopkins2012,oklopcic2017}. Indeed, simulation work by \cite{buck2017} incorporating more moderate feedback models find little evidence for clumps of stars after 200 Myr. Studies where feedback effects are modeled as radiation pressure and supernovae, however, predict massive clumps ($>10^{8}\Msun$). We find that only a small fraction of the clump masses (3 of 15; before disk-subtraction) observed in G04-1 are consistent with this mass regime.  While the present paper provides data for only a single galaxy, at least in this case the majority of clumps (both before and after disk-subtraction) observed in G04-1 appear to fall in the range $10^{6-8}\Msun$, which is fairly near the middle of the mass spectrum observed in high-redshift observations of lensed galaxies and in simulations.  

\cite{lenkic2021} recently studied the internal color gradients of clumps in DYNAMO galaxies. They find that color gradients of clumps are more commonly consistent with changes in age, rather than extinction. This is different than the explanation by \cite{buck2017} that clumps in turbulent galaxies are the result of regions of lower extinction. They also find that DYNAMO clump age gradients are consistent with an old clump of stars, with a young center. This is consistent with our results, in which old stellar clumps co-exist with high SFR surface densities. Together, these results give two independent lines of evidence of DYNAMO galaxies contain long-lived clumps of stars.

The total mass of the clumps (after local disk subtraction) is approximately 3.1$\times 10^{8}\Msun$. Using high resolution \textit{HST} continuum data, \citealt{fisher2017a} fit a surface brightness profile for G04-1 and estimate a bulge-to-total ratio of about 11\%. Given a total stellar mass of 6.47$\times 10^{10}\Msun$, this corresponds to a current bulge mass of about 7.1$\times 10^{9}\Msun$. If we assume that all of the clumps observed in G04-1 survive sufficiently long to migrate inward, then the total contribution of these clumps to the mass of the bulge would be of about 5\%. Assuming there is no further mass growth of the galaxy, this potential contribution increases to 8.54$\times 10^{8}\Msun$ or about 12\% when considering non-disk subtracted mass estimates. Whether or not the current observations lend support to the notion of bulge building from clump infall clearly depends on the duty cycle of the process, because a single episode of bulge growth from this process would only add incrementally to the present bulge.

\subsection{Placing Clump Masses and Sizes in Context}

A number of recent observational studies have shown that the star forming clumps observed in $1<z<3$ galaxies may not be as massive or large as initially predicted due to observational challenges inherent to mapping light in galaxies at high redshift. For example, studies examining systems with strong gravitational lensing have been able to explore clumps at spatial resolutions below 100 pc (e.g. \citealt{livermore2012,livermore2015,wuyts2014}) and consistently find smaller clump sizes ranging 50 pc - 1 kpc. To explore the effect of resolution on derived clump properties, \citet{cava2018} evaluated multiple images of a single galaxy with different lensing magnifications (consistent with effective resolutions ranging from 30 to 300 pc) and found that at coarse resolution clump sizes and masses were systematically overestimated. Their target galaxy, the Cosmic Snake, has similar total mass and SFR as ours in this work. They find that when observed at the finest spatial resolution that clumps in the Cosmic Snake galaxy have masses ranging $\sim10^{7}-10^{8.5}$~M$_{\odot}$, a range quite comparable to ours. Using machine learning methods to identify clumps in VELA zoom-in simulations, \citealt{huertas2020} find observational effects significantly impact clump properties, leading to a factor of 10 over-estimation in stellar mass. While it is becoming increasingly clear that observational constraints and resolution limits have led to significant ambiguity in the true masses and sizes of clumps, it's difficult to ascertain to what degree these biases have impacted the values we report here for clumps in G04-1. In this section, we aim to contextualize our work by comparing our findings with similar studies spatially resolving star formation in galaxies.

Using a combination of \textit{HST} broad and narrow band imaging, \citet{bastian2005} explore star cluster populations in M51. M51 can be taken as a typical star-forming spiral disks, and is therefore useful for comparison to the extreme star formation in G04-1. They measure masses in complexes of star formation with sizes from $\sim100$~pc to a few hundred parcecs, comparable to the sizes in G04-1. The key difference is the stellar masses associated to the star forming associations in M51 are $3-30\ \times\ 10^{4}\ \Msun$, which is multiple orders of magnitude lower than what we observe in G04-1.

Ultra luminous infrared galaxies (U/LIRGs) are a class of local ($z<0.1$) dusty systems with high IR luminosities ($L_{[8-1000\mu m]}>10^{11}\Lsun$) and associated SFRs which are comparable to our target and to galaxies at $z\sim 1-3$ (of order $\rm 10^{2-3}\ \Msun\ yr^{-1}$) \citep{malek2017,larson2020}.  Looking at resolved star formation (via Pa $\alpha$ and Pa $\beta$ line emission) in 48 galaxies from the Great Observatories All-Sky LIRG Survey (GOALS) with \textit{HST}, \citet{larson2020} find that while the typical sizes of clumps in LIRGs are comparable to that observed in DYNAMO ($\sim100-900$ pc), they appear generally less massive ($M_{*,median}\sim 5 \times 10^{5}\ \Msun$) and exhibit individual SFRs which are roughly an order of magnitude lower ($\sim0.03\ \Msun \rm yr^{-1}$).

\citet{messa2019} examined star forming clumps in 14 galaxies from the Lyman-Alpha Reference Sample (LARS; $z=0.03-0.2$) via UV/optical data from \textit{HST}. They find that for LARS galaxies (which are selected to be similar to high-$z$ galaxies based on their $\rm H\alpha$ and UV fluxes) clump sizes range from $20-600$ pc with a median clump size of about $d\sim60$ pc. While these reported clump sizes are smaller than that seen in DYNAMO ($<$R$_{clump}>\ \sim$500 pc; \citealt{fisher2017a}) it remains unclear whether this indicates a significant difference in clump sizes or if it is due to the clustering and resolution effects discussed above. For example, the smallest clump sizes reported by \citet{messa2019} are observed in the lowest redshift LARS galaxies where the best resolution of their data ($\sim 10$ pc) is roughly 10$\times$ better than in the \textit{HST} imaging presented in \S2.3 for G04-1.

Global system dynamics are closely linked with theoretical formation pathways for star forming clumps in galaxies and thus also provide a critical point of comparison \citep[discussion in][]{fisher2017b}. Like other DYNAMO galaxies, G04-1 has been shown to have a well ordered rotation field measured both in ionized gas at AO-enabled high spatial resolution \citep{oliva2018} and in stellar kinematics \citep{bassett2014}. Moreover, \cite{fisher2017a} shows that the star-light profile is consistent with an exponential model. 
While some fraction of ULIRGs are likely spirals, the primary dynamical driving mechanism of the bulk of U/LIRGs is widely considered to be merging (e.g. Larson et al. 2016). This has been shown to be the true case of LIRG galaxies in GOALS, where the sample is dominated by interacting systems \citep{larson2020}. We highlight this important kinematic distinction as it presents an important caveat when comparing the observed properties of clumps in DYNAMO with other local high SFR samples.

\begin{figure}[htb!]
\begin{center}
\includegraphics[scale=0.52]{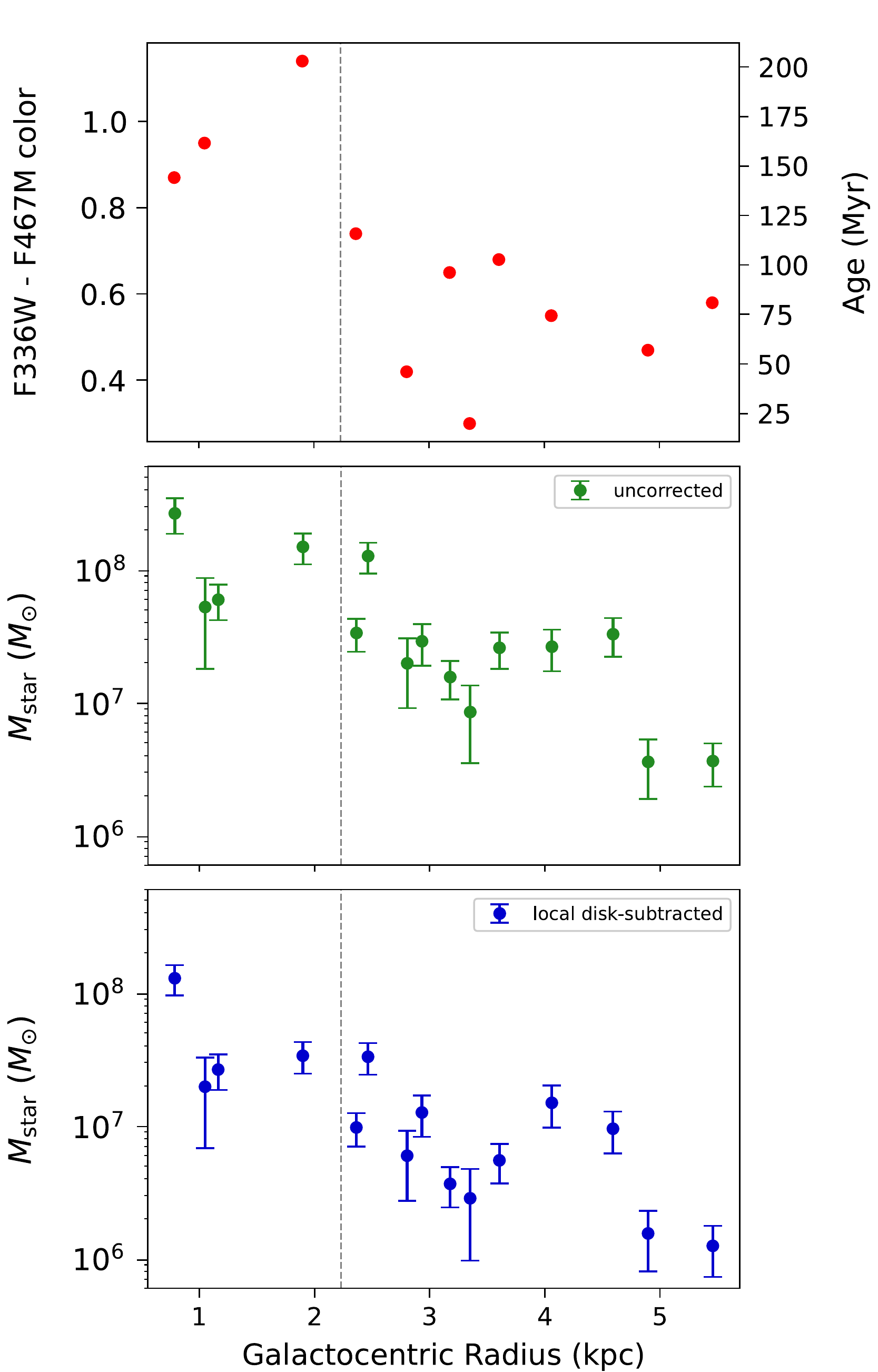}
\end{center}
\caption{The relationship between clump \textit{HST} F336W - F467M (un-disk corrected) color (top panel, solid red circles), mass (middle panel; green circles), disk-subtracted mass (bottom panel; blue circles) and galactocentric radius. Here, green points represent mass values uncorrected for disk contribution and dark blue points represent mass estimates for which we have subtracted off the local disk background. Radial distance values (horizontal axis) are calculated as the offset between the individual clump aperture centroid and the galaxy center. The Pearson's correlation coefficients we calculate for these three quantities indicate strong negative correlation: we find $r$-values of -0.65, -0.64, and -0.61 for the clump colors, uncorrected masses, and disk-corrected masses, respectively. The vertical grey dashed line represents the outer edge of the ring region in G04-1 at $R_{GC}\sim2.2$ kpc.}
\label{fig:mass_vs_dist}
\end{figure}

\subsection{Radial Gradients in Clump Properties}
In Fig. 6 we show the radial gradients in both clump stellar masses and clump colors. We find that clumps inside $R<2$ kpc have average disk-corrected masses of $5.2\times 10^{7}\Msun$ and average non-disk corrected F336W-F467M colors of $\sim$0.8. Where clumps outside of this radius appear smaller and bluer with masses $9.1\times 10^{6}\Msun$ and F336W-F467M values of about $\sim$0.6. The trend with mass is roughly consistent with a log-linear decline in mass with radius. The observable properties of clumps: e.g. the number, mass, star formation rate (SFR), and age, are fundamental for testing feedback models. In the case of G04-1, there is clear evidence for a clump mass gradient across the galaxy. We have computed Pearson's correlation coefficients for the three derived quantities in Fig. \ref{fig:mass_vs_dist} and find $r$-values of -0.65, -0.64, and -0.61 for the clump colors, uncorrected masses, and disk-corrected masses, respectively. This indicates that all three quantities exhibit strong negative gradients across the galaxy disk. This is shown in Fig. \ref{fig:mass_vs_dist} (bottom panel) which illustrates the radial dependence of clump mass (both before and after disk-subtraction). Clumps closer to the galaxy's nucleus are observed to be significantly (more than a factor of ten) more massive than those at the outskirts. We note that galaxy G04-1 is host to various structural features, including multiple spiral arms and a prominent nuclear ring, which are clearly observed in the top row and central panels in Fig. \ref{fig:clump_apertures}.  This ring structure appears somewhat asymmetric and varies radially in width due to the existence of bright knots of ongoing star formation and (in some part) the slight inclination of the galaxy. We have estimated the radial outer edge of this nuclear ring region to be located at a distance of roughly $R_{GC}\sim 2.2$ kpc (the average value of four measurements made at cardinal points in the \textit{HST} $\rm H\alpha$ ratio map image) from the galaxy centre. For spatial reference, we include a vertical dashed line denoting the outer edge of this ring region in the panels of Fig. \ref{fig:mass_vs_dist}.

This clump mass gradient in G04-1 could have multiple origins. In particular, it could be due to (\textit{i}) the inward migration of clumps while gradually forming more stars, (\textit{ii}) an inside-out growth of the galaxy disk, or (\textit{iii}) the Jeans mass being larger at smaller radii. From spatially-resolved kinematic maps (using Keck-OSIRIS observations of the Pa-$\alpha$ line; see \citealt{oliva2018}) it is known that the gas velocity dispersion in G04-1 declines slightly with radius. While this would argue for a higher Jeans mass near the galaxy centre, the declining surface density and age gradient make this scenario (\textit{iii}) unlikely. From Fig. \ref{fig:mass_vs_dist}, it is observed that from $r=5$ kpc to $r=1$ kpc in the disk, clump age increases by about 150 Myr while the overall clump mass increases by about $\sim 10^{8} \Msun$. \citet{lenkic2021} performs multi-band stellar population modelling of the spectral energy distributions of clumps in G04-1 and finds a similar range in clump ages ($80 - 300$ Myr). If this variation is entirely due to star formation (e.g. scenario \textit{i}), this would imply that the mean SFR of clumps during migration should be $\sim 0.7 \Msun yr^{-1}$. This value is surprisingly consistent with the average clump SFR for this galaxy reported by \cite{fisher2017a} using H$\alpha$-based measurements ($<$SFR$\rm_{clump}>$ $\sim 0.83\ \Msun\ yr^{-1}$). This also suggests that an upper-limit to possible mass contributed by clumps in G04-1 to the bulge (assuming all clumps complete migration) is of order $10^{9} \Msun$ or addition of roughly $\sim 14\%$ to the bulge mass. The SFR of clumps is likely a more reliable proxy for mass growth than mass flow rate, as more mass is likely to be lost via feedback than star formation. Notably, the 150 Myr age-span observed for clumps in G04-1 is well-matched to the minimum age predicted for clumps which are able to survive long enough to complete in-spiral to the galaxy center \citep{guo2012,guo2017,fs2011,shibuya2016,soto2017}. However, we qualify this simple model by noting that it assumes that 1) clumps originate in the outskirts of the disk, 2) clumps are isolated from other clumps (i.e. they don't merge) and 3) all clumps within the galaxy survive long enough to complete their migration inward. It remains unclear to what degree these assumptions are reasonable.

We find no clear evidence of a radial dependence on the number density of clumps; Fig. 1 illustrates that stellar clumps are indeed evenly distributed across the disk. Additionally, Fig. \ref{fig:mass_vs_dist} (top panel) shows that stellar clumps near the nuclear ring appear (albeit to a small degree) consistently more red, suggesting that these stellar populations may be slightly older and more evolved.  \cite{fisher2017a} report star formation rates (SFR) for these clumps (\S2, and listed in Table \ref{tab:prop}) and find higher SFRs for clumps located near the inner ring (on average 1.6 $\Msun$/yr) when compared to the arm (0.5 $\Msun$/yr).

Our measurements appear to be consistent with observations of clump properties in high-redshift galaxies, as well as with numerical investigations. In their simulations, \cite{mandelker2014,mandelker2017} and \cite{dekel2021} find significant gradients in clump properties across the disk. More specifically, clumps closer to the galaxy center tend to be both more massive and comprised of older stellar populations (i.e. longer-lived). 
Fig. \ref{fig:mass_vs_dist} is consistent with the investigation clump properties at high-redshift by \cite{guo2018}, who examined UV-bright clumps in CANDELS galaxies and found significant gradients in both stellar mass (with inner clumps on average being more massive than outer clumps by 1-2 orders of magnitude) and color. They also found that inner clumps appear redder (in $U-V$) than those observed in the outskirts of the disks. \cite{fs2011} performed deep \textit{HST} (NIC2/F160W and ACS/F814W) imaging of clumps in six $z\sim 2$ star forming galaxies and also found central clumps to be more massive and older. Positive color and stellar mass gradients were similarly observed by \cite{cava2018} in imaging of clumps within the `Cosmic Snake', a lensed system. Similar mass-radius relations also appear at low redshift: observations of star clusters in local galaxies such as $z<0.1$ star forming spirals \citep{sun2016} and major merger Arp 299 \citep{randriamanakoto2019} find cluster mass increases with decreasing galactocentric radius. Radial trends are often inferred to be an observational basis for clump migration. If the clumps in G04-1 are long-lived, however, and have survived long enough for in-spiral to establish these gradients, then the masses we observe strongly argue against very strong feedback effects (see \S1).

It is emphasized that the clump mass results shown in Fig. \ref{fig:mass_vs_dist} are quite robust; these radial trends are certainly not due to uncertainties in mass-to-light calculations, since our observations use K-band imaging where M/L is very stable. Moreover, the high spatial resolution in two dimensions allows for an accurate disk subtraction and, therefore, radial trends cannot be due to background subtraction effects.

\begin{figure}[htb!]
\begin{center}
\includegraphics[scale=0.55]{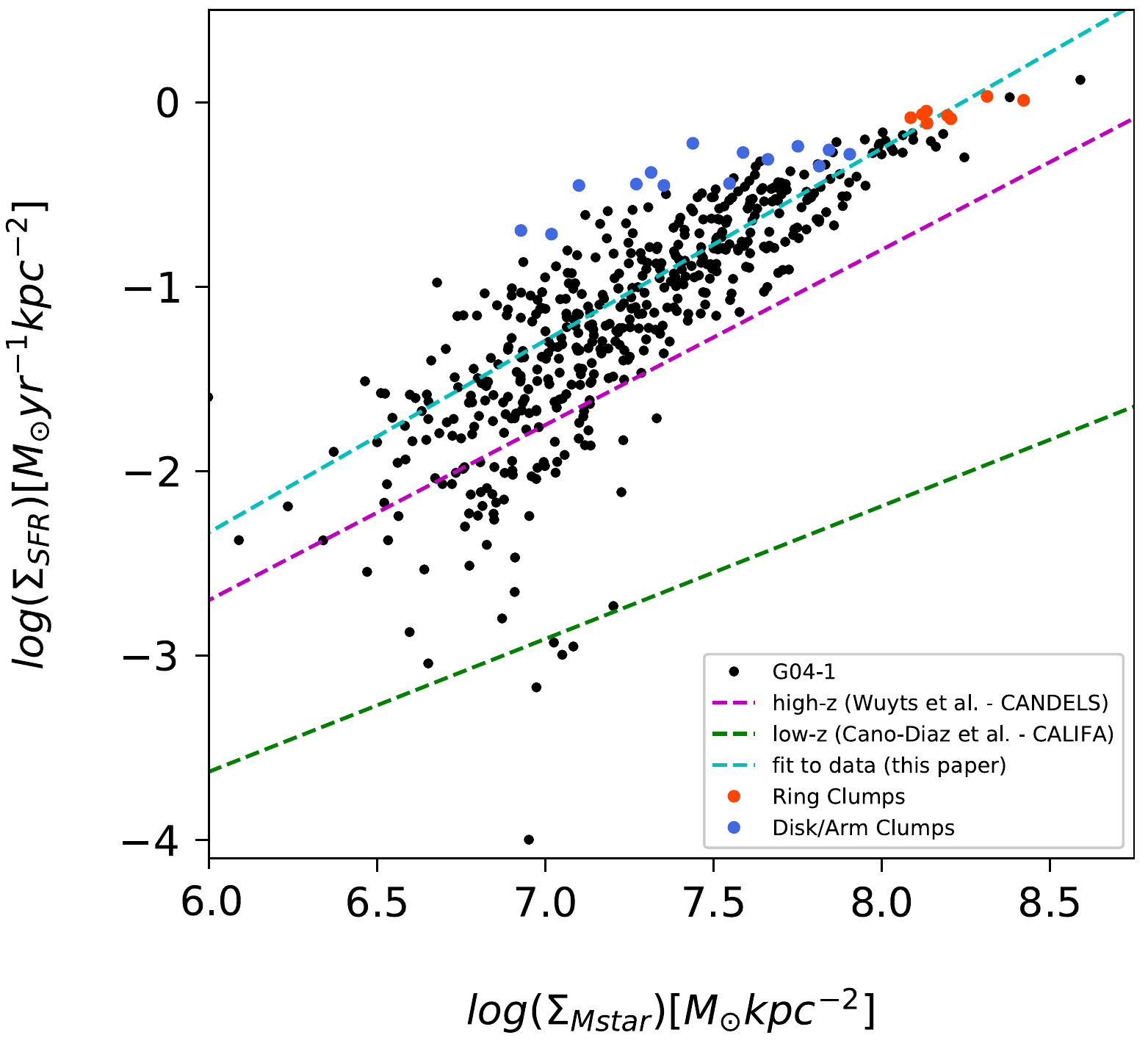}
\end{center}
\caption{The resolved star forming main-sequence for G04-1. Here we show the relationship between star formation rate (taken from \citealt{fisher2017a}) and stellar mass (from this work) for $\sim$400pc sized regions in G04-1. Galaxy regions which can be directly associated with clumps (identified in our NIRC2 K-band maps) are identified in this figure by being either red or blue, depending on whether they are positioned in the nuclear ring or disk regions of the galaxy. The power-law fit to the data presented in this paper (described in \S 4.2) is shown as a dashed cyan line. For comparison, we include similar relations derived from observations at high-z \citep{wuyts2013} and low-z \citep{cd2016}.}
\label{fig:SFR_vs_Mstar}
\end{figure}

\subsection{Resolved $\Sigma_{SFR}$ versus $\Sigma_{star}$}
Observations at both high and low redshift have found a tight ($\sim$0.3 dex) and roughly linear relationship (the so-called ``star forming main-sequence") between the star formation rate and stellar mass surface densities in actively star forming galaxies \citep{daddi2007,noeske2007,elbaz2011}. In Fig. \ref{fig:SFR_vs_Mstar}, we plot the star forming main-sequence in terms of spatially-resolved quantities for G04-1. To estimate the SFR and stellar mass surface densities ($\Sigma_{SFR}\ \&\ \Sigma_{M_{star}}$, respectively) both the $K_{P}$-band (this work) and the H$\alpha$ maps (presented in \cite{fisher2017a}) were re-gridded to a common scale of $\sim$400 pc/pixel. New fluxes were calculated for each pixel in each of the re-gridded maps and used to estimate corresponding SFRs and stellar mass surface density values. A galaxy-averaged K-band mass-to-light ratio of 0.1 (corresponding to the integrated F336W - F467M color for the galaxy) was used to calculate stellar masses. As a reference, all pixels associated with the stellar clump regions are plotted as red or blue points (depending on whether they correspond to ring or disk/arm clumps; see Fig. \ref{fig:SFR_vs_Mstar}). All non-clump region pixels (defined as where $<$30\% of the contributing flux originates from clumps) are shown as black circles. We note that the two non-clump data points in Fig. \ref{fig:SFR_vs_Mstar} with high (ring-like) $\Sigma_{SFR}$ values correspond to regions near the centre of G04-1.

After flagging the clump regions as red and blue in Fig. \ref{fig:SFR_vs_Mstar}, we note the following: (1) the clump regions within the galaxy define the upper end of SFR surface density for this galaxy and (2) support for the radial trends in clump properties described in \S4.1 at larger spatial-scales. Large-sample resolved studies of galaxies (see \citealt{wuyts2013,hemmati2014,cd2016,magdis2016}) find that this spatially-resolved relation is observed across redshifts and wavelengths, but the derived slopes and zero-points vary. A power-law fit to all of the data points in Fig. \ref{fig:SFR_vs_Mstar} results in a slope that is remarkably near unity: 

\begin{equation}
\rm log(\Sigma_{SFR}) = -8.58\pm0.21\ +\ 1.041\pm0.03\log(\Sigma_{M_{star}}). 
\end{equation}

\noindent A separate fit of only the data points associated with clumps results in a flatter relation: 
\begin{equation}
\rm log(\Sigma_{SFR}) = -3.69\pm3.81\ +\ 0.44\pm0.49\ log(\Sigma_{M_{star}}).
\end{equation}

\noindent We note that when computing these relations, we omitted a number of very low signal-to-noise data points at large radii in both the H$\alpha$ and $K_{P}$ maps, because they were sufficiently near the sky background their errors were likely dominated by systematics from sky subtraction.

\cite{cd2016} use integral field spectroscopy (IFS) observations of 306 local galaxies from the CALIFA survey ($0.005<z<0.03$) to derive this relation and find $\rm log(\Sigma_{SFR}) = -7.95 + 0.72\ log(\Sigma_{M_{*}})$. This is significantly less steep than that observed at high-$z$: \cite{wuyts2013} use kpc-scale multi-wavelength broad-band imaging (from CANDELS) and H$\alpha$ surface brightness profiles (from 3D-HST) for 473 star forming galaxies ($0.7<z<1.5$) and find $\rm log(\Sigma_{SFR}) = -8.4 + 0.95\ log(\Sigma_{M_{*}})$. For reference, we plot both of these observed relations in Fig. \ref{fig:SFR_vs_Mstar}.

Nearly all regions within G04-1 directly overlap with the star forming main-sequence relation derived from observations of high-$z$ galaxies. Indeed, the slope and intercept values are remarkably similar. In terms of its integrated properties, G04-1 lies offset from the star forming main-sequence. However, its location at $z\sim 0.1298$ results in some ambiguity as to the reason for this offset. For example, one wonders whether G04-1 is a normal, local star forming galaxy which simply hosts high-SFR clumps (i.e. a scenario where the clumps were completely externally formed and then accreted).  From this figure we infer that this is likely not to be the case. Instead, Fig. \ref{fig:SFR_vs_Mstar} suggests that all regions (both the clump and intra-clump regions) within the galaxy are experiencing an enhanced mode of star formation, more like what is routinely observed in galaxies at high-$z$. 

\subsection{Stellar clumps are co-located with Ha}
Where possible, we calculated the spatial offsets between the locations of the NIR clumps discussed in the present paper and their counterparts in the \cite{fisher2017a} H$\alpha$ maps. On average, clumps in K-band have centers (determined via centroiding; see \S3) which are displaced from their corresponding H$\alpha$ centers (from \citealt{fisher2017a}) by about 2.6 pixels ($\sim$0.1$\arcsec$). This close alignment of clumps can be visualized in the enlarged clump multi-band panels provided in Fig. \ref{fig:clump_apertures}. As this average offset between clumps is very similar to the width of our night's AO PSF ($\sim$3 pix), we cannot infer whether this offset is indeed real or a manifestation of the PSF associated with our night of data. Clumps in the ring region of G04-1 appear more offset from their H$\alpha$ counterpart than those in the galaxy's arms. However, as stated in \S3, a number of the apertures (12, 13, \& 14) taken from \cite{fisher2017b} for clumps in the ring region of the galaxy required transformation (rotation and aperture-size modification) in order to adequately encompass the K-band clump flux. Indeed, we identify fewer clumps in the ring of G04-1 than \cite{fisher2017a}. These differences are a likely consequence of the significant amount of additional light contributed from the central disk component in the K-band imaging. This light may be washing out the concentrated light from individual clumps within the ring and blending structure. These effects would present reasonable explanations for ring clumps exhibiting greater offsets. 

In general, the stellar clumps appear well-aligned with the active star forming regions observed in the H$\alpha$ map, implying that these more evolved stellar populations maintain a link with regions of recent star formation. If stellar clumps are long-lived structures then this would suggest they don't undergo a single burst and then shut off, but that they continue to experience star formation. However, we observe a number of clumps (IDs: 4, 7, 11, \& 15) for which we do not observe an obvious H$\alpha$ component. This is quite interesting because, as seen in Fig. 5, we observe that all of the regions in G04-1 associated with clumps exhibit high observed star formation rate surface densities. This would imply that all stellar clumps should be H$\alpha$-bright. We have identified two possible scenarios which may explain this: the light from these HII regions may be obscured by the existence of a molecular cloud situated along our line-of-sight. While this would certainly be a very specific situation, it is statistically possible as typical HII regions are of order a few tens of parsecs and only a few would be required to produce something of order the scale of a clump. An alternative scenario would be that these clumps have indeed turned off in terms of their star formation and are now simply wandering through the gas rich disk. This second scenario is quite interesting as numerical simulations (e.g. \citealt{bournaud2014}) suggest that clumps massive enough to survive as long-lived should eventually re-accrete gas from the disk and re-ignite in star formation. Certainly, more work teasing out the details of these stellar populations is required to determine which of these scenarios is more likely.


\section{Summary}\label{sec:5}
In this paper, we present a case-study of stellar clumps in a gas-rich, clumpy turbulent disk galaxy from the DYNAMO sample.
\begin{itemize}
    \item We present new K-band imaging of G04-1 using Keck-NIRC2 and \textit{Hubble Space Telescope} WFC/ACS observations using the F336W ($U$) and F467M (Stromgen $b$) filters.
    \item We identify 15 clumps in K-band light of G04-1 that are evenly distributed in mass, ranging from 0.36 to $27.0\times10^{7}\ \Msun$. These values correspond to Clump IDs 14 \& 9, respectively. Subtraction of the local disk component from clump light results in a drop in clump mass estimates of around 50\%. This corresponds to a median disk-corrected clump mass of $\sim 1.5\times 10^{7}\ \Msun$.
    \item We find evidence of radial trends in clump stellar properties. Clumps closer to the galaxy nucleus are observed to be more massive and appear consistently more red, suggesting that these stellar populations may be more evolved. We do not find evidence of a radial dependence on the number density of clumps.
    \item We investigate the relationship between the star formation rate and stellar mass surface densities using high-resolution maps in $K_{P}$ (from in this paper) and H$\alpha$ (from \textit{HST}; presented in \citealt{fisher2017b}). A power-law fit to the data results in slope and intercept values (1.041 $\pm$ 0.03 \& -8.58 $\pm$ 0.21, respectively) similar to that derived from populations of high-$z$ galaxies. Indeed, nearly all regions in G04-1 appear to be undergoing an enhanced mode of star formation.
\end{itemize}

\section*{Acknowledgements}
Some of the data presented herein were obtained at the W. M. Keck Observatory, which is operated as a scientific partnership among the California Institute of Technology, the University of California and the National Aeronautics and Space Administration. The Observatory was made possible by the generous financial support of the W. M. Keck Foundation. 

The authors wish to recognize and acknowledge the very significant cultural role and reverence that the summit of Maunakea has always had within the indigenous Hawaiian community.  We are most fortunate to have the opportunity to conduct observations from this mountain.

The above work is also partly based on observations made with the NASA/ESA \textit{Hubble Space Telescope}, obtained from the Data Archive at the Space Telescope Science Institute, which is operated by the Association of Universities for Research in Astronomy, Inc., under NASA contract NAS 5-26555. 

This publication makes use of data products from the Two Micron All Sky Survey (2MASS), which is a joint project of the University of Massachusetts and the Infrared Processing and Analysis Center/California Institute of Technology, funded by the National Aeronautics and Space Administration and the National Science Foundation.

This research made use of \textsc{Astropy}, a community-developed core Python package for Astronomy (Astropy Collaboration, 2013).

HAW and RGA thank NSERC and the Dunlap Institute for Astronomy and Astrophysics for financial support. The Dunlap Institute is funded through an endowment established by the David Dunlap family and the University of Toronto.

DBF acknowledges support from Australian Research Council (ARC) Future Fellowship FT170100376 and ARC Discovery Program grant DP160102235. ADB acknowledges partial support from AST1412419.

KG acknowledges support from Australian Research Council (ARC) Discovery Program (DP) grant DP130101460. Support for this project is provided in part by the Victorian Department of State Development, Business and Innovation through the Victorian International Research Scholarship (VIRS).

\bibliographystyle{apj}
\bibliography{dynamo}
\newpage
\end{document}